\documentclass[]{raa}            

\usepackage{graphicx,times}  
\usepackage{natbib}
\usepackage{amssymb,amsmath}
\usepackage{amsmath}
\usepackage{multirow}
\usepackage{tabularx}
\usepackage{booktabs}
\usepackage[flushleft]{threeparttable}
\usepackage{lipsum}
\usepackage{enumitem}
\usepackage{hyperref}
\usepackage{siunitx}
\usepackage{rotating}
\usepackage{adjustbox} 
\usepackage{longtable}
\usepackage{xltabular}
\usepackage{xcolor}
\usepackage{textcomp}
\bibpunct{(}{)}{;}{a}{}{,}

\usepackage{tikz,xcolor,hyperref}

\definecolor{lime}{HTML}{A6CE39}
\DeclareRobustCommand{\orcidicon}{%
	\begin{tikzpicture}
		\draw[lime, fill=lime] (0,0) 
		circle [radius=0.16] 
		node[white] {{\fontfamily{qag}\selectfont \tiny ID}};
		\draw[white, fill=white] (-0.0625,0.095) 
		circle [radius=0.007];
	\end{tikzpicture}
	\hspace{-2mm}
}

\foreach \x in {A, ..., Z}{%
	\expandafter\xdef\csname orcid\x\endcsname{\noexpand\href{https://orcid.org/\csname orcidauthor\x\endcsname}{\noexpand\orcidicon}}
}

\begin{document}
	
	\title{Prediction of short stellar activity cycles using derived and established empirical relations between activity and rotation periods.
	}
	
	\volnopage{Vol.0 (20xx) No.0, 000--000}      
	\setcounter{page}{1}          
	
	\author{A. k. Althukair
		\inst{1,2}\orcidA{}
		\and D. Tsiklauri
		\inst{1}\orcidB{}
	}

	\institute{Department of Physics and Astronomy, School of Physical and Chemical Sciences, Queen Mary University of London,
		Mile End Road, London, E1 4NS,
		UK; {\it a.k.althukair@qmul.ac.uk}, {\it d.tsiklauri@qmul.ac.uk}\\
		\and
		Physics Department, College of Sciences, Princess Nourah Bint Abdulrahman University, Riyadh, PO Box 84428, Saudi Arabia\\
		\vs\no
		{\small Received 20xx month day; accepted 20xx month day}}
	
	\abstract{ In our previous work, we investigated the occurrence rate of super-flares on various types of stars and their statistical properties, with a particular focus on G-type dwarfs, using entire Kepler data. The said study also considered how the statistics change with stellar rotation period, which in turn, had to be determined. Using such new data, as a by-product, we found 138 Kepler IDs of F and G types main sequence stars with rotation periods less than a day ($P_{\rm rot}<1$ d). On one hand, previous studies have revealed short activity cycles in  F-type and G-type stars and the question investigated was whether or not short-term activity cycles are a common phenomenon in these stars. On the other hand, extensive studies exist which establish empirical connection between a star's activity cycle and rotation periods. In this study, we compile all available Kepler data with $P_{\rm rot}<1$ d, derive and use, established empirical relation between $P_{\rm cyc}$ and  $P_{\rm rot}$ with the aim to provide predictions for very short $5.09\leq P_{\rm cyc}\leq 38.46$ d cases in a tabular form. We propose an observation to measure $P_{\rm cyc}$ using monitoring program of stellar activity (e.g. activity-related chromospheric emission S-index) or similar means for the Kepler IDs found in this study in order put to test the derived here empirical relations between $P_{\rm cyc}$ and  $P_{\rm rot}$. We also propose an alternative method for measuring very short $P_{\rm cyc}$, using flare-detection algorithms applied to future space mission data.
	}
	\authorrunning{Althukair \& Tsiklauri }            
	\titlerunning{Prediction of short stellar activity cycles}  
	
	\maketitle
	
	%
	%
	\section{Introduction}\label{section:introduction3}
	The 11-year cycle of solar activity discovered by Schwabe in 1844 \citep{schwabe1844}, is a significant phenomenon in solar and stellar physics. The cycle is manifested by a periodic change in solar activity, including the appearance of sunspots and changes in the Sun's magnetic field
	on this time-scale. Smoothed sunspot numbers have been widely used as a proxy for solar activity over the past four centuries \citep{Shepherd2014}. 
	The idea of the sunspot number was first introduced by \citet{1961says.book.....W} in the mid-19th century, and it has since become a standard measure for quantifying solar activity. These numbers reveal that there are almost regular cycles of about 11 years, reflecting the Sun's magnetic activity. 
	
	During the course of a solar cycle, the Sun experiences alternating periods of strong and weak activity known as solar maximum and minimum \citep{Hathaway2002,Shepherd2014,Reinhold2017}. As the solar cycle progresses, the magnetic field becomes more complex and twisted. This results in the emergence of sunspots, which are dark areas on the surface of the Sun with intense magnetic fields, vary in size and can last from days to several months \citep{Petrovay1997}, decaying into bright areas called faculae formed by smaller magnetic concentrations \citep{Reinhold2017}. During the active phase of the solar cycle (solar maximum), {the sunspots number increases, and their size becomes larger on the surface of the Sun.} At the same time, bright faculae also become more prominent. As the cycle progresses, the number of sunspots decreases, the overall brightness of the Sun {reduces}, and the Sun enters its least active phase of the solar cycle (solar minimum). These dark and bright features on the Sun's surface contribute to the variability in the total solar irradiance (TSI) \citep{Marchenko2022}. Therefore, the TSI data can capture the combined effects of the evolving dark and bright features during the solar cycle \citep{Domingo2009,Reinhold2017}.
	
	Cyclic activity has been observed in stars other than the Sun through long-term brightness changes associated with increased occurrence of active regions on their surfaces or in their lower stellar atmospheres \citet{Reinhold2017}. The Mount Wilson HK program, which started in 1966 and lasted until the end of the 20th century, was the first to conduct a systematic search for activity cycles in main sequence stars \citep{Wilson1978,Baliunas1995,Mittag2019}. {By examining chromospheric emission within the Ca II H\&K spectral lines, the magnetic field associated with active regions on stellar surfaces is vital in conveying energy to the chromosphere. This heightened influx of energy into the chromosphere results in amplified chromospheric emission, notably observable in the central regions of the Ca II H\&K spectral lines, as indicated by \citet{Reinhold2017}.} The measure of the chromospheric emission strength is described by the Mount Wilson S-index \citep{Vaughan1978} or by the quantity $R^{\prime}_{\rm HK}$ \citep{Brandenburg2017}. \citet{Vaughan.Preston1980} investigated the chromospheric activity levels in main-sequence F-G-K-M stars by measuring the chromospheric CaII H\&K emission fluxes. They noted that these stars display varying degrees of chromospheric activity and observed a noticeable lack in the number of F-G stars displaying intermediate activity compared to both highly active and less active stars. They suggested that the absence of such stars could be attributed to a decline in chromospheric activity as the stars age. \citet{Noyes1984a} examined the relationship between chromospheric activity, specifically the $R^{\prime}_{\rm HK}$ activity index, and the Rossby number Ro = $P_{\rm rot}/\tau_{\rm c}$ for a sample of main-sequence stars of spectral type F or later, where $P_{\rm rot}$ is the rotational period of the star and $\tau_{\rm c}$ is a theoretically derived convective turnover time. They found a strong correlation between the $R^{\prime}_{\rm HK}$ activity index and the Rossby number. However, in contrast to the findings of \citet{Vaughan.Preston1980}, \citet{Noyes1984a} did not find any signs of the "Vaughan-Preston gap". \citet{Noyes1984b} investigated the empirical relation between rotation period $P_{\rm rot}$, spectral type, and activity cycle period $P_{\rm cyc}$ for 13 slowly rotating main-sequence stars. They found that the cycle period is related to the rotation period by a power law: $P_{\rm cyc}\propto P_{\rm rot}$$^{\rm 1.25}$. This relationship can alternatively be expressed as
	$P_{\rm cyc}\approx$ Ro$^{1.25} \approx (P_{\rm rot}/\tau_{\rm c})$$^{1.25}$ \citep{Brandenburg2017,Mittag2023}. For stars of spectral type G0-K5, \citet{Baliunas1995} observed a pattern of variation in the rotation period and the measure of chromospheric activity (S-index). Their research revealed that the chromospheric activity levels were high in young stars with fast rotation periods. Chromospheric activity and rotation rates of stars in the intermediate age range were average. Alternatively, the chromospheric activity levels were low in old stars with slow rotation periods. This observation supports the existence of the Vaughan-Preston gap, indicating that chromospheric activity and rotation change over time as the stars age. The relation between rotation periods and activity cycles of a sample of stars was investigated by \citet{Baliunas1996}, who discovered a correlation between the two variables. In particular, they observed that stars with slower rotation periods exhibit longer activity cycles, while stars with faster rotation periods tend to have shorter activity cycles. According to \citet{Olah2002}, the relation between rotation periods and cycle lengths is more evident for stars with shorter activity cycles. However, the association becomes less clear for longer cycle lengths when considering more recent findings on the time variability of solar cycles. 
	
	{In order to provide background of results in this field of research we now discuss previous literature.}
	\citet{Vida2013} investigated the behaviour and activity cycles of four fast-rotating late-type stars with ($P_{\rm rot}\leq 0.5$ days), highlighting the presence of 1-year cycles and the correlation between rotation rate and cycle length. \citet{Vida2014} used the short-term Fourier transform, a time-frequency analysis method, to examine the light curves of 39 fast-rotating late-type active stars with rotation periods of less than one day. Nine of the selected stars showed indications of activity cycles with periods between 300 and 900 days.  {These cycles were inferred based on the observed variations in the typical latitude of the starspots. These variations, along with the differential rotation of the stellar surface, result in changes in the observed rotation period during the activity cycle.} This variation in the rotation period was attributed to the movement and evolution of starspots at different latitudes of the star. 
	
	\citet{Reinhold2017} used four years of Kepler data to determine the cyclic variations in the amplitude of the light curve and the rotation period of stars by analysing a sample of active stars and calculating the rotation period and variability amplitude for each star in each Kepler quarter. Then they searched for periodic variations in these time series using Lomb-Scargle periodograms and employed a false alarm probability (FAP) criterion for selection. The study's findings indicate that amplitude periodicities, associated with underlying activity cycles, are detected in 3203 stars with cycle periods ranging from 0.5 to 6 years and rotation periods ranging from 1 to 40 days. According to \citet{Brandenburg2017} analysis of new observations and previous data, the longer and shorter cycle periods closely match expectations based on the average activity levels and rotation periods, which indicates a connection between stellar activity and stellar rotation. 
	
	\citet{Baliunas1995} reported an activity cycle of 11.6 years in the F-type star $\tau$ Boo (HD 120136). However, the authors assigned a FAP "poor" grade to this finding. \citet{Mittag2017} detected an activity cycle with a duration of 122 days in their analysis of the S-index data of $\tau$ Boo. This short activity cycle periods suggest that $\tau$ Boo may exhibit variations on a relatively short timescale. \citet{Mittag2019} focused on exploring the presence of short-term activity cycles in F-type stars, specifically using S-index time series data obtained with the TIGRE telescope. They utilized the generalized Lomb-Scargle periodogram method to analyze the data and search for periodic variations with a maximum length of 2 years. Their sample of F-type stars identified four stars that exhibited cyclic variations with periods of less than a year. However, compared to solar-type stars with well-developed cyclic activity, the amplitude of these short-term cyclic variations in F-type stars was smaller. Based on their findings, \citet{Mittag2019} concluded that the activity behaviour among F-type stars differs from that of the Sun and cooler main sequence stars. 
	
	By studying 44 main-sequence stars with confirmed activity cycles, and rotation periods, \citet{Mittag2023} examined the relation between the length of the activity cycle and the Rossby number (Ro). They used empirical turnover periods based on the B-V colour index to calculate Rossby numbers, from which they deduced an empirical relationship between the Rossby number and the cycle duration. The study showed linear behaviour in the double-logarithmic relationship between the Rossby number and cycle period. In addition, the relative convection zone depth was found to be correlated with cycle length and convective turnover time.
	
	{Besides the 11-year solar cycle, shorter cycles were discovered called the Rieger cycles. The original Rieger cycles were first identified in the Sun by \citet{Rieger1984} with a specific periodicity of approximately 154 days for flare occurrences. The Rieger-type cycles (RTCs) encompass cycles with periods (PRTC) ranging from 109 to 276 days. These RTCs were observed in various phenomena beyond solar flares, such as solar magnetic field and sunspot indexes, indicating their widespread nature. The underlying nature of RTCs remains unclear. The RTCs become more pronounced during the solar activity maximum. There is a potential connection between RTCs and the modulation of the solar magnetic dynamo process, as discussed in \citet{Arkhypov2021} and references therein. 
Possible reasons encompass the role of inertial g- and r-waves, also known as Rossby waves, as modulators of the emergence of magnetic flux in the Sun.  \citet{Arkhypov2021} analyze photometric data from 1726 main-sequence stars with varying effective temperatures and rotation periods to study RTCs in other stars.
		Two types of RTCs are identified among the surveyed stars. The activity cycles with RTC periods ($P_{\rm RTC}$) are independent of the stellar rotation period and are suggested to be driven by Kelvin waves. The second type are activity cycles with $P_{\rm RTC}$ proportional to the stellar rotation period and are suggested to be driven by Rossby waves.}
	
	\citet{Parker1955} model of the $\alpha \textendash \Omega$ dynamo introduced the concept of migratory dynamo waves, which play a crucial role in generating the observed solar cycle \citep{Mittag2023}. The $\alpha \textendash$effect, arising from the twisting of rising magnetic field tubes due to Coriolis forces, creates the poloidal magnetic field required for the next sunspot cycle. This effect is responsible for the reversal of magnetic polarities between successive cycles \citep{Parker1955,Mittag2023}. On the other hand, the $\Omega \textendash$effect, resulting from the differential rotation of the star, generates a toroidal magnetic field by stretching the magnetic field lines in a longitudinal direction. The combination of the $\alpha \textendash$effect and the $\Omega \textendash$effect leads to the formation of migratory dynamo waves, where the toroidal field is periodically regenerated and transformed into the poloidal field through the action of the $\alpha \textendash$effect. These migratory dynamo waves propagate and interact within the star's convective zone, causing the cyclic variations in the magnetic field \citep{Mittag2023}.
	
	{Now we describe existing theoretical knowledge about the possible relation between the magnetic cycle period and the rotation period of a star. In this context,} according to \citet{Noyes1984b}, 
	the magnetic cycle period for G and K dwarfs with convective turnover times ($\tau_{\rm c}$) between 11 and 26 days, is found to be proportional to the rotation period as follows:
	\begin{equation}
		1/P_{\rm cyc} \propto(\tau_{\rm c} / P_{\rm rot})^{n},
		\label{eq1}
	\end{equation}
	where n is 1.25.
	
	{Simple dynamo models were discussed for understanding stellar magnetic activity and their implications for magnetic cycle periods in stars. \citet{Stix1981} derived an equation to determine the critical dynamo number $D_{\rm crit}$ given by
		\begin{equation}
			D_{\rm crit}\sim (\Omega \tau_{c})^{2} \times\left(\frac{R_{\star}}{l}\right)^{3},
			\label{D_crit}
		\end{equation}
		where $(\Omega \tau_{c})^{2}$ is the inverse squared Rossby number and $R_{\star}/l$ is the relative depth of turbulence. suggesting that the occurrence of dynamo action is contingent upon the interplay between stellar rotation and stellar structure.}
	
	{\citet{Parker1955} provides relation for the magnetic cycle frequency, $\omega_{\rm mag\_cyc}$, that involves the shear, $H$, and the $\alpha$-effect. \citet{Stix1976}  presented an equivalent expression for the magnetic cycle frequency derived by \citet{Parker1955}, in terms of angular velocity gradient, $\Omega^{\prime}$, given by
		\begin{equation}
			\omega_{\rm mag\_cyc}= |\alpha \Omega^{\prime}|^{\frac{1}{2}},
			\label{mag-cyc}
		\end{equation}
		indicating a proportional relationship between the cycle frequency and rotation frequency. Based on the model's assumptions equation \ref{mag-cyc} can be written as }
	
	\begin{equation}
		P_{\rm mag\_cyc}=2 P_{\rm cyc}\approx\sqrt{\dfrac{R_{\star}}{l}}P_{\rm rot}.
		\label{eq2}
	\end{equation}
	{where $l$ here is the length scale of turbulence and $R_{\star}$ is the stellar radius.
		This equation indicate the theoretical prediction of the relation between star's activity cycle and its rotation periods, which is equation (6) in \citet{Mittag2023}.} 
	
	According to the simple theoretical arguments quoted by \citet{Mittag2023},
	the magnetic cycle period $P_{\rm mag\_cyc}$ is proportional to the rotation period $P_{\rm rot}$. However, there is a modifying factor, $l/R_{\star}$ the relative depth of turbulence, which depends on the stellar structure, which itself may depend on the effective temperature or B-V color index of the star. {This factor is expected to vary among different stars, especially those with different sizes, masses, and ages. The smallness of the inverse relative depth of the turbulence ensures that the period of the magnetic activity cycle $P_{\rm mag\_cyc}$ is small.
However, precisely what factors guarantee smallness of $R_{\star}/l$
in a particular star is poorly understood. That is why it is unclear
why stars with very short activity cycles, studied in this paper, exist. 
All we can surmise is that the above theoretical arguments suggest 
$P_{\rm mag\_cyc}$ should scale as $\propto \sqrt{{R_{\star}}/{l}}$.} 
	
	{Activity cycles, characterised by variations in magnetic activity over time, are essential for understanding the fundamental mechanisms that drive the magnetic fields of stars. A range of methodologies exist for the identification of activity cycles in stars. One such approach involves integrated flux measurements by continuously monitoring the total amount of energy emitted by a star, enabling the detection of variations in its magnetic activity \citep{Kopp2016,Reinhold2020}. Another approach is the analysis of chromospheric emission lines from the outer atmosphere of a star. Additionally, tracking of starspots by observing the movement and changes in starspots on a star's surface serves as an indicator for fluctuations in magnetic activity \citep{Montet2017}.
		Nevertheless, it is important to acknowledge that these techniques include limitations regarding photometric precision and the small sample size in spectroscopic observations \citep{Scoggins2019}. An alternative approach involves detecting flares. Flares are a frequently observed phenomenon resulting from magnetic activity and are easier to detect, even at significant distances from stars. Wide-field photometric surveys allow for simultaneous monitoring of stars, making it possible to survey them for flare activity \citep{Scoggins2019}. The Sun experiences variations in its flare rate by a factor of 10 between the solar maximum and minimum activity periods. \citet{Scoggins2019} focused on fluctuations in the frequency of flares from stars detected by the Kepler mission. The study examined a sample of 347 flare stars, which were selected based on having measured Kepler rotation periods, a minimum of 100 candidate flare events. \citet{Scoggins2019} aimed to identify coherent variations in flare activity among these stars by computing the fractional luminosity emitted in flares. One star, KIC 8507979, was identified as the best candidate for flare activity variation. This star has a rotation period of 1.2 days and emits an average of 0.82 flares per day with energies exceeding $10^{32}$ erg over the 18 Kepler quarters. The study observed a decline in flare activity from KIC 8507979 over time. Although the flare census derived from the Kepler light curve of this star did not provide definitive evidence for a stellar activity cycle, the observed variation of approximately 0.1 dex per year was consistent with cyclic behavior over ten years or more.}
	
	{The motivation for the current work is as follows:} In paper I \citep{paper1}, we looked for super-flares on different types of stars and focused on G-type dwarfs using entire Kepler data to study 
	various aspects of statistical properties of the occurrence rate of super-flares.
	In paper II \citep{paper2},  as a by-product, we found thirteen peculiar Kepler IDs that are Sun-like, slowly rotating with rotation periods of 24.5 to 44
	days, and yet can produce a super-flare and  six G-type and four M-type Kepler IDs with exceptionally large amplitude super-flares. As noted previously, 
	these detections defy our current understanding of stars and hence deserve a further investigation.
	In this paper III, the last in this series, {we use the same data set as in \citet{paper1} in order to study} empirical connection between a star's activity cycle and rotation periods for a sample of F and G main sequence stars with rotation periods of less than one day. 
	Here our aim is to provide predictions for very short activity cycle cases in a tabular form and to investigate in the future whether these short activity cycles are a common phenomenon in these stars or not.  Section \ref{section:Method3} presents the method used in this work which includes the reproduction of \citet{Mittag2023} fit, the data representation and fit and the target selection method. The main findings of the study are presented in Section \ref{section:results3}, and section \ref{section:conclusion3} concludes this work with our main conclusions.
		
	\section{Methods}\label{section:Method3}
	In our study, we adopt the terminology used by \citet{Brandenburg2017,Mittag2023} to categorize branches into two types: the "inactive" branch, referred to as the short-cycle branch $P_{\rm cyc}^{S}$ and the "active" branch, referred to as the long-cycle branch $P_{\rm cyc}^{L}$. These terms were introduced first time in \citet{Brandenburg2017}. According to \citet{Mittag2023} this notation is more accurate and aligned with the actual characteristics of the branches. Therefore, they suggested that these terms should be used in future studies to refer to the two branches.
	
	\subsection{Reproduction of \citet{Mittag2023} $P_{\rm cyc}^{S}$ vs. $P_{\rm rot}$ Fit}\label{subsection:Mittag}
\begin{figure}[h!]
		\centering
		\includegraphics[width=0.6\textwidth]{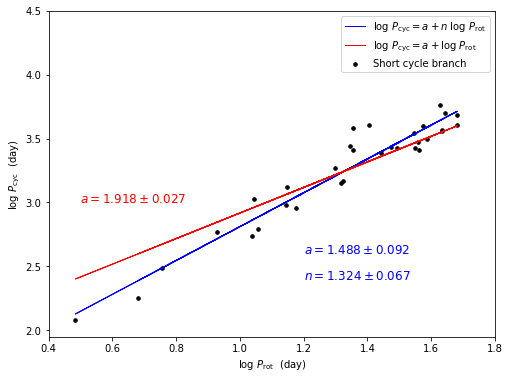}
		\caption{Log-scale of rotation period versus log-scale of {observed activity} cycle period (short cycle branch) for a sample of stars taken from \citet{Mittag2023}. The deduced fit of $P_{\rm rot}$ vs. $P_{\rm cyc}$ relation are shown as solid lines. The blue line shows the fit when slope $n$ is treated as an independent parameter while the red line shows the fit with a fixed slope of $n$=1.}
		\label{Mittag-fit}
	\end{figure}

	In this subsection, we reproduced the fit between $P_{\rm cyc}^{S}$ and $P_{\rm rot}$ data from \citet{Mittag2023} to derive the fit parameters. First, we collected the data in Table\ref{tab:long}, the first 32 rows {are the observed activity cycle on the short-cycle branch $P_{\rm cyc}^{S}$ from \citet{Mittag2023} Table 1,} along with the 32 corresponding rotation periods $P_{\rm rot}$. These cycle lengths and rotation periods can be found in Table 1. Then we plotted in logarithmic scale the rotation periods on the x-axis versus the calculated cycle period on the y-axis as shown in Figure \ref{Mittag-fit}, using the empirical relation in \citet{Mittag2023} between the cycle periods and rotation periods in logarithmic terms that is given by:
	\begin{equation}
		\log P_{\rm cyc}\approx a+n \log P_{\rm rot}.
		\label{eq3}
	\end{equation}
	Since the theoretical relation, equation \ref{eq2} 
	implies a linear connection between $P_{\rm cyc}$ and $P_{\rm rot}$, we fitted the data using Python \textit{least-square} fit, a common technique for determining the best-fitting parameters for a given model, for two different slope adjustments as in \citet{Mittag2023}. Also, we computed the $R^{2}$ \textit{coefficient of determination} to measure how well the model fits the data. A $R^{2}$ value of 1 means that the predictions from the regression fit the data perfectly. First, we set the slope $n$ to be 1 and deduced the value of $a$ parameter as {$a$ = 1.918 $\pm$ 0.027 and the value of $R^{2}$= 0.87}. The red line in Figure \ref{Mittag-fit} illustrates this trend. Then we repeated the fit by treating slope $n$ as an independent variable to derive $a$ and $n$ values as equation \ref{eq3} now becomes:
	{
		\begin{equation}
			\log P_{\rm cyc}\approx (1.488 \pm 0.092)+(1.324 \pm 0.067) \log P_{\rm rot}.
			\label{eq4}
		\end{equation}
	}
	
	 {and the value of $R^{2}$= 0.93.} The blue line in Figure \ref{Mittag-fit} represents this fit. It is obvious that the $n$ = 1 relation does not fit the short periods data, as \citet{Mittag2023} pointed out.
	
	\subsection{Data representation and fit}
	In this subsection, we repeat the fit between $P_{\rm rot}$ and $P_{\rm cyc}^{S}$ using a larger data sample,  taken from {\citet{Mittag2023}} and other previous studies.
	
	This sample, shown in Table\ref{tab:long}, contains {92} $P_{\rm rot}$ and their {92} corresponding $P_{\rm cyc}^{S}$. {In addition to the 32 observed activity cycles in \citet{Mittag2023}, we aggregated 32 activity cycles on the short-cycle branch $P_{\rm cyc}^{S}$ computed by \citet{Mittag2023} together with the corresponding 32 rotation periods $P_{\rm rot}$. Furthermore, we included 28 activity cycles and their corresponding rotation periods that were collected from various other studies. Thses $P_{\rm cyc}^{S}$ were taken from \citet{Lopes2015,Egeland2015,BoroSaikia2016,Flores2016,Brandenburg2017,Mittag2019}}
	The star ID, effective temperature ($T_{\rm eff}$), color index (B-V), $P_{\rm rot}$ and $P_{\rm cyc}$ are shown in Table\ref{tab:long}. Unavailable data is left blank in the table. It should be noted that
	{we used in the fit two $P_{\rm cyc}^{S}$ values for each of the 32 stars IDs in \citet{Mittag2023}, one was the observed $P_{\rm cyc}^{S}$ by a previous study, and the other was the calculated $P_{\rm cyc}^{S}$ by \citet{Mittag2023}.} Except for HD 16673 {for which we collected three $P_{\rm cyc}^{S}$} due to the multiple sources, as shown in Table \ref{tab:long}. References for each $P_{\rm rot}$ and $P_{\rm cyc}^{S}$ are shown in Table \ref{tab:long}.
	
\begin{figure}[h!]
		\centering
		\includegraphics[width=0.6\textwidth]{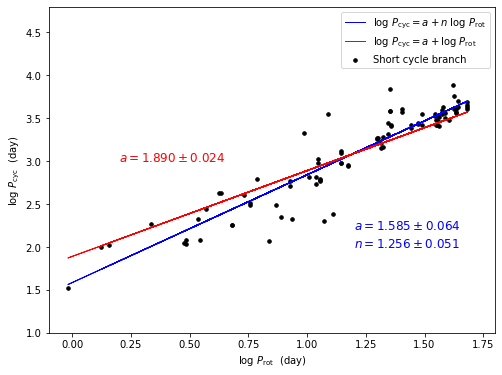}
		\caption{Log-scale of rotation period versus log-scale of cycle period (short cycle branch) for a {92} samples of stars taken from previous studies in Table\ref{tab:long}. The deduced fit of $P_{\ rot}$ vs. $P_{\ cyc}$ relation are shown as solid lines. The blue line shows the fit where slope $n$ is treated as an independent parameter while the red line shows the fit with a fixed slope of $n$=1.}
		\label{all-Pcyc-fit}
	\end{figure}

	In the same way as in subsection \ref{subsection:Mittag}, we used the empirical relation between $P_{\rm rot}$ and $P_{\rm cyc}$ in logarithmic scale given by equation \ref{eq3} using the new data set in Table\ref{tab:long} to produce the fit parameters $a$ and $n$. We performed a \textit{least-square} fit in Python to fit the data using two different slope adjustments again, one with a fixed slope $n$ of 1 and another with the $n$ treated as a free variable. This fit is shown in Figure \ref{all-Pcyc-fit}.  For the fit with a fixed slope of 1, we determined the value for the parameter {$a$= 1.890 $\pm$ 0.024} and $R^{2}$= 0.83. This trend is shown by the red line in Figure \ref{all-Pcyc-fit}. While for the fit with the slope $n$ treated as a free variable, we deduced values for the parameters $a$ and $n$ as {$a$=1.585 $\pm$ 0.064, $n$=1.256 $\pm$ 0.051} and $R^{2}$= 0.87. This fit is represented by the blue line in Figure \ref{all-Pcyc-fit}. So that equation\ref{eq3} becomes now
	{
		\begin{equation}
			\log P_{\rm cyc}\approx (1.585 \pm 0.064)+(1.256 \pm 0.051) \log P_{\rm rot}.
			\label{eq5}
		\end{equation}
	}
	
	We note that our value of {$n=1.256 \pm 0.051$} with the extended dataset is 
	closer to \citet{Noyes1984b}'s $n=1.25$ than \citet{Mittag2023}'s $n= 1.324 \pm 0.067$.
	
	\begin{xltabular}{\textwidth}{|ccc|cc|cc|}
		\caption{list of star IDs with their parameters, used in previous studies.} \label{tab:long} \\
		
		\hline \multicolumn{1}{|c}{\rm HD/KIC} & \multicolumn{1}{c}{$T_{\rm  eff}$} & \multicolumn{1}{c}{\rm B-V} & \multicolumn{1}{|c}{$P_{\rm  rot}$\rm [d]} &\multicolumn{1}{c|}{\rm Ref}&\multicolumn{1}{c}{$P_{cyc}^{S}$\rm [yr]}&\multicolumn{1}{c|}{Ref}\\ \hline \hline
		\endfirsthead
		
		\multicolumn{7}{c}%
		{\tablename\ \thetable{} -- continued from previous page} \\
		\hline
		\multicolumn{1}{|c}{\rm HD/KIC} & \multicolumn{1}{c}{$T_{\rm  eff}$} & \multicolumn{1}{c}{\rm B-V} &\multicolumn{1}{|c}{$P_{\rm  rot}$\rm [d]} &\multicolumn{1}{c|}{\rm Ref}&\multicolumn{1}{c}{$P_{cyc}^{S}$\rm [yr]}&\multicolumn{1}{c|}{Ref}\\ \hline \hline
		\endhead
		
		\hline \multicolumn{7}{|c|}{{Continued on next page}} \\ \hline
		\endfoot
		
		\hline
		\hline
		\endlastfoot
		Sun&5777&0.642&25.4$\pm$1&1&11$\pm$2&1\\
		HD 3651&5211&0.85&44&1&13.8$\pm$0.4&1\\
		HD 4628&5120&0.89&38.5$\pm$2.1&1&8.6$\pm$0.1&1\\
		HD 10476&5244&0.836&35.2$\pm$1.6&1&9.6$\pm$0.1&1\\
		HD 10780&5321&0.804&22.14$\pm$0.55&2&7.53$\pm$0.16&2\\
		HD 16160&5060&0.918&48$\pm$4.7&1&13.2$\pm$0.2&1\\
		HD 16673&6183&0.524&5.7&3&0.847$\pm$0.006&5\\
		HD 17051&6045&0.561&8.5$\pm$0.1&1&1.6&1\\
		HD 22049&5140&0.881&11.1$\pm$0.1&1&2.9$\pm$0.1&1\\
		HD 26965&5282&0.82&43&1&10.1$\pm$0.1&1\\
		HD 30495&5804&0.632&11.4$\pm$0.2&1&1.7$\pm$0.3&1\\
		HD 32147&4801&1.049&48&1&11.1$\pm$0.2&1\\
		HD 43587&5876&0.61&22.6$\pm$1.9&4&10.44$\pm$3.03&4\\
		HD 75332&6089&0.549&4.8&5&0.493$\pm$0.003&5\\
		HD 75732&5167&0.869&37.4$\pm$0.5&6&10.9&13\\
		HD 76151&5714&0.661&15&1&2.5$\pm$0.1&1\\
		HD 100180&6013&0.57&14&1&3.6$\pm$0.1&1\\
		HD 103095&5449&0.754&31&1&7.3$\pm$0.1&1\\
		HD 120136&6245&0.508&3.05$\pm$0.01&7&0.333$\pm$0.002&7\\
		HD 128621&5098&0.9&36.2$\pm$1.4&1&8.1$\pm$0.2&1\\
		HD 140538&5645&0.684&20.71$\pm$0.32&8&3.88$\pm$0.02&8\\
		HD 146233&5741&0.652&22.7$\pm$0.5&1&7.1&1\\
		HD 149661&5265&0.827&21.1$\pm$1.4&1&4$\pm$0.1&1\\
		HD 160346&4975&0.959&36.4$\pm$1.2&1&7$\pm$0.1&1\\
		HD 165341 A&5188&0.86&19.9&1&5.1$\pm$0.1&1\\
		HD 166620&5151&0.876&42.4$\pm$3.7&1&15.8$\pm$0.3&1\\
		HD 185144&5366&0.786&27.7$\pm$0.77&2&6.66$\pm$0.05&2\\
		HD 190406&5910&0.6&13.9$\pm$1.5&1&2.6$\pm$0.1&1\\
		HD 201091&4764&1.069&35.4$\pm$9.2&1&7.3$\pm$0.1&1\\
		HD 219834 B&5055&0.92&43&1&10$\pm$0.2&1\\
		KIC 8006161&5234&0.84&29.8$\pm$3.1&1&7.4$\pm$1.2&1\\
		KIC 10644253&5943&0.59&10.9$\pm$0.9&1&1.5$\pm$0.1&1\\
		\hline
		Sun&5777&0.642&25.4$\pm$1&1&10.3&14\\
		HD 3651&5211&0.85&44&1&11.7&14\\
		HD 4628&5120&0.89&38.5$\pm$2.1&1&9.9&14\\
		HD 10476&5244&0.836&35.2$\pm$1.6&1&9.2&14\\
		HD 10780&5321&0.804&22.14$\pm$0.55&2&5.6&14\\
		HD 16160&5060&0.918&48$\pm$4.7&1&12.4&14\\
		HD 16673&6183&0.524&5.7&3&0.9&15\\
		HD 17051&6045&0.561&8.5$\pm$0.1&1&1.4&14\\
		HD 22049&5140&0.881&11.1$\pm$0.1&1&2.6&14\\
		HD 26965&5282&0.82&43&1&11.5&15\\
		HD 30495&5804&0.632&11.4$\pm$0.2&1&1.6&14\\
		HD 32147&4801&1.049&48&1&11.7&15\\
		HD 43587&5876&0.61&22.6$\pm$1.9&4&10.4&14\\
		HD 75332&6089&0.549&4.8&5&0.5&15\\
		HD 75732&5167&0.869&37.4$\pm$0.5&6&9.7&14\\
		HD 76151&5714&0.661&15&1&2.4&14\\
		HD 100180&6013&0.57&14&1&3.4&14\\
		HD 103095&5449&0.754&31&1&9.6&14\\
		HD 120136&6245&0.508&3.05$\pm$0.01&7&0.3&14\\
		HD 128621&5098&0.9&36.2$\pm$1.4&1&9.2&14\\
		HD 140538&5645&0.684&20.71$\pm$0.32&8&4.5&14\\
		HD 146233&5741&0.652&22.7$\pm$0.5&1&7.2&14\\
		HD 149661&5265&0.827&21.1$\pm$1.4&1&5.3&14\\
		HD 160346&4975&0.959&36.4$\pm$1.2&1&9&14\\
		HD 165341 A&5188&0.86&19.9&1&4.9&14\\
		HD 166620&5151&0.876&42.4$\pm$3.7&1&11.1&14\\
		HD 185144&5366&0.786&27.7$\pm$0.77&2&7.3&14\\
		HD 190406&5910&0.6&13.9$\pm$1.5&1&2.6&14\\
		HD 201091&4764&1.069&35.4$\pm$9.2&1&8.3&14\\
		HD 219834 B&5055&0.92&43&1&11&15\\
		KIC 8006161&5234&0.84&29.8$\pm$3.1&1&7.7&14\\
		KIC 10644253&5943&0.59&10.9$\pm$0.9&1&1.8&14\\
		\hline
		102712791& &0.277&0.96$\pm$0.03&9&0.09$\pm$0.008&9\\
		102720703& &0.514&10.2$\pm$0.6&9&1.781$\pm$0.356&9\\
		102721955& &0.431&2.17$\pm$0.06&9&0.512$\pm$0.055&9\\
		102723038& &1.404&8.6$\pm$0.5&9&0.575$\pm$0.019&9\\
		102726103& &0.767&3.7$\pm$0.1&9&0.759$\pm$0.058&9\\
		102738457& &0.592&12.9$\pm$0.6&9&0.655$\pm$0.06&9\\
		102749950& &0.657&5.4$\pm$0.2&9&1.118$\pm$0.071&9\\
		102750723& &1.143&1.44$\pm$0.02&9&0.29$\pm$0.019&9\\
		102754736& &0.48&6.9$\pm$0.3&9&0.321$\pm$0.022&9\\
		102758108& &0.641&6.1$\pm$0.2&9&1.682$\pm$0.151&9\\
		102770332& &2.055&4.2$\pm$0.1&9&1.162$\pm$0.112&9\\
		102770893& &0.874&4.3$\pm$0.2&9&1.17$\pm$0.123&9\\
		102777006& &1.177&1.33$\pm$0.02&9&0.277$\pm$0.022&9\\
		102778595& &1.157&11.8$\pm$0.7&9&0.551$\pm$0.041&9\\
		102780281& &1.304&3$\pm$0.1&9&0.301$\pm$0.022&9\\
		61 Cygni A ( HD 201091)&4545&1.069&35.7$\pm$1.9&10&7.2$\pm$1.3&10\\
		HD 100563& & &7.73&5&0.61&5\\
		HD 114710&5970&0.58&12.3$\pm$1.1&1&9.6$\pm$0.3&1\\
		HD 128620&5809&0.71&22.5$\pm$5.9&1&19.2$\pm$0.7&1\\
		HD 16673&6183&0.524&7.4$\pm$0.07&5&0.85&5\\
		HD 201092&4040&1.37&37.8$\pm$7.4&1&11.7$\pm$0.4&1\\
		HD 219834 A &5461 &0.8 & 42&1&21&1\\
		HD 49933& & &3.45&5&0.58&5\\
		HD 78366&5915&0.63&9.7$\pm$0.6&1&5.9$\pm$0.1&1\\
		HD 81809&5623&0.8&40.2$\pm$3&1&8.2$\pm$0.1&1\\
		solar analog HD 30495&5826&0.632&11.36$\pm$0.17&11&1.67$\pm$0.35&11\\
		solar analog HD 45184&5871&0.62&19.98$\pm$0.02&12&5.14&12\\
		$\tau$ Boo&&0.48&3.5&5&0.33&5\\
	\end{xltabular}
	\begin{minipage}{14cm}
		\small  {\textbf{Notes:} The table illustrates a list of stars ID with their corresponding $\rm B\textendash V$ values, effective temperature $T_{\rm  eff}$, the rotation period $P_{\rm rot}$ with the reference number and the short branch cycle period $P_{\rm cyc}^{S}$ with the reference number.\\
			\textbf{References:} (1) \citet{Brandenburg2017}, (2) \citet{Olspert2018}, (3) \citet{Noyes1984b}, (4) \citet{Ferreira2020}, (5) \citet{Mittag2019}, (6) \citet{Mittag2017a}, (7) \citet{Mittag2017}, (8) \citet{Mittag2019b}, (9) \citet{Lopes2015}, (10) \citet{BoroSaikia2016}, (11) \citet{Egeland2015}, (12) \citet{Flores2016} , (13) \citet{Baum2022}, (14) \citet{Mittag2023}.}
	\end{minipage}
	
	\subsection{Data Samples}\label{section:Data3}
	One of the main challenges in studying the relation between cycle length and rotation period is the lack number of well-known and accurately measured activity cycles. This limitation introduces uncertainties in the derived empirical relations \citet{Mittag2023}. To overcome these challenges, it is crucial to obtain more reliable cycle periods, particularly for long-period cycles. Achieving this requires long-term time series observations of stars to gather comprehensive and accurate data on their activity cycles \citet{Mittag2023}. Therefore, when looking for activity cycles, it is more efficient to monitor fast-rotating objects, as cycles can be discovered within a few years of observation, as opposed to stars with longer rotation periods \citet{Vida2013}. For this reason, we chose our sample for this study to include fast-rotating main-sequence stars of type F and G from Kepler data with well-known rotation periods of less than one day. First, we collected all Kepler IDs which has well-known rotation periods. We then selected targets with rotation periods of less than a day. Using \textit{Gaia} Data Release 2 (\textit{Gaia}-DR2), we identified F- and G-type main sequence stars by their effective temperatures and radius based on the Harvard Spectral classification. The ranges of the effective temperature are 6000-7500 K and 5200-6000 K for F and G types, respectively. We thus obtained a total of 811 Kepler IDs of F- and G- type stars with less than one day rotation period. By using the radius restriction of the main-sequence stars as 1.15-1.4 $\rm R_{\odot}$ and 0.96-1.15 $\rm R_{\odot}$ for F and G types, respectively, the final data sample reduced to 138  Kepler targets with a number of 83 F-type and 55 G-type main-sequence stars. 71.74\% of the rotation periods for these stars were taken from \citet{McQuillan2014}. 15.94\% from \citet{Santos2021}, 5.07\% from \citet{Reinhold2015}, 4.35\% from \citet{Chowdhury2018} and 2.90\% from \citet{Yang2019}. These 138 Kepler targets are listed in Table \ref{tab:less than 1d} with their effective temperature, radius, rotation period and the references for these rotation periods.
	
	\section{Results}\label{section:results3}
	
	Using a data set of 138 Kepler IDs with $P_{\rm rot}$ ranging from 0.202 d to 0.997 d, we provide a
	prediction for the corresponding value of their $P_{\rm cyc}^{S}$, by applying the empirical relation between $P_{\rm cyc}$ and $P_{\rm rot}$ with the derived parameters in equation \ref{eq5}. Hence we 
	obtained the predicted values of $P_{\rm cyc}$ from 
	{
		\begin{equation}
			P_{\rm cyc}\approx 10^{(1.585 \pm 0.064)+(1.256 \pm 0.051) \log P_{\rm rot}}.
			\label{eq6}
		\end{equation}
	}

	From  equation \ref{eq6}, we calculated 138 $P_{\rm cyc}$ for 83 F-type and 55 G-type main-sequence stars whose rotation period is less than a day. The shortest $P_{\rm cyc}$ is equal to {5.09 d}  while the longest $P_{\rm cyc}$ is equal to {38.46 d}. All the 138 predicted $P_{\rm cyc}$ are listed in Table \ref{tab:less than 1d}
	
	\begin{xltabular}{\textwidth}{|cccccc|cccccc|}
		\caption{lists of the 138 Kepler IDs with their parameters and predicted $P_{\rm cyc}$.} \label{tab:less than 1d} \\
		
		\hline \multicolumn{1}{|c}{\rm KIC} &\multicolumn{1}{c}{$T_{\rm  eff}$} & \multicolumn{1}{c}{$R_{\odot}$} & \multicolumn{1}{c}{$P_{\rm  rot}$\rm [d]}&\multicolumn{1}{c}{\rm Ref} &\multicolumn{1}{c|}{$P_{\rm cyc}$\rm [d]}&\multicolumn{1}{c}{\rm KIC} & \multicolumn{1}{c}{$T_{\rm  eff}$}& \multicolumn{1}{c}{$R_{\odot}$} & \multicolumn{1}{c}{$P_{\rm  rot}$\rm [d]}&\multicolumn{1}{c}{\rm Ref} &\multicolumn{1}{c|}{$P_{\rm cyc}$\rm [d]} \\ \hline \hline
		\endfirsthead
		
		\multicolumn{12}{c}%
		{\tablename\ \thetable{} -- continued from previous page} \\
		\hline
		\multicolumn{1}{|c}{\rm KIC} &\multicolumn{1}{c}{$T_{\rm  eff}$} & \multicolumn{1}{c}{$R_{\odot}$} & \multicolumn{1}{c}{$P_{\rm  rot}$\rm [d]}&\multicolumn{1}{c}{\rm Ref} &\multicolumn{1}{c|}{$P_{\rm cyc}$\rm [d]}&\multicolumn{1}{c}{\rm KIC} & \multicolumn{1}{c}{$T_{\rm  eff}$}& \multicolumn{1}{c}{\rm $R_{\odot}$} & \multicolumn{1}{c}{$P_{\rm  rot}$\rm [d]}&\multicolumn{1}{c}{\rm Ref} &\multicolumn{1}{c|}{$P_{\rm cyc}$\rm [d]} \\ \hline \hline
		\endhead
		
		\hline \multicolumn{12}{|r|}{{Continued on next page}} \\ \hline
		\endfoot
		
		\hline
		\endlastfoot
		
		757099&5521&1.05&0.36&1&10.66&6877871&6508&1.40&0.54&2&17.74\\
		1028018&5544&1.14&0.62&2&21.10&6948098&6095&1.29&0.57&3&18.98\\
		1721795&6534&1.31&0.89&2&33.22&6961285&5802&0.98&0.45&2&14.11\\
		1872192&5316&0.98&0.67&2&23.26&6962901&5601&0.97&0.98&2&37.50\\
		2557335&5568&1.01&0.24&2&6.41&7199002&6381&1.24&0.57&2&18.98\\
		2558273&6673&1.35&0.99&2&37.98&7199013&5286&0.96&0.57&2&18.98\\
		2715228&6374&1.30&0.99&1&37.98&7199037&6024&1.36&0.57&2&18.98\\
		2715410&5997&1.11&0.90&1&33.69&7354297&5481&1.05&0.95&2&36.06\\
		2849645&5424&1.06&1.00&2&38.46&7461022&6168&1.28&0.59&2&19.82\\
		2985825&6783&1.23&0.94&3&35.58&7678509&6644&1.22&0.96&2&36.54\\
		3124412&6302&1.21&0.93&1&35.11&7707736&5644&1.09&0.76&2&27.25\\
		3241517&6283&1.34&0.78&3&28.15&7816211&6050&1.32&0.29&2&8.12\\
		3352959&6476&1.37&0.76&2&27.25&7909399&6574&1.40&0.82&2&29.97\\
		3356577&6746&1.39&0.63&4&21.53&7915824&6231&1.39&0.74&2&26.35\\
		3448722&5872&1.13&0.41&2&12.55&7973882&5512&1.06&0.35&2&10.29\\
		3448817&6792&1.33&0.95&4&36.06&8016369&6734&1.34&0.77&1&27.70\\
		3459311&5789&1.05&0.98&2&37.50&8043256&6680&1.27&0.93&2&35.11\\
		3550386&6006&1.30&0.32&2&9.19&8144578&6639&1.32&0.59&2&19.82\\
		3836772&6210&1.32&0.69&2&24.13&8197275&5604&1.14&0.44&2&13.71\\
		3869099&5607&1.01&0.29&2&8.12&8264155&6738&1.33&0.91&4&34.16\\
		4175618&5369&1.05&0.41&2&12.55&8264659&5417&1.12&0.97&1&37.02\\
		4283120&6202&1.25&0.52&2&16.92&8285970&5639&1.14&0.57&2&18.98\\
		4374659&5824&1.03&0.23&2&6.07&8313378&6624&1.31&0.54&2&17.74\\
		4386947&5681&1.14&0.65&2&22.39&8382253&5695&1.01&0.63&3&21.53\\
		4464528&6392&1.38&0.22&2&5.74&8393626&5893&1.15&0.43&2&13.32\\
		4464530&6545&1.30&0.22&2&5.74&8420730&5770&1.08&0.25&2&6.74\\
		4570231&5661&0.99&0.54&1&17.74&8651921&6473&1.29&0.95&2&36.06\\
		4660562&5677&0.96&0.77&1&27.70&8687209&5650&1.00&0.77&1&27.70\\
		4762130&6202&1.35&0.80&2&29.06&8804962&6586&1.23&0.90&2&33.69\\
		4774370&6546&1.36&0.93&2&35.11&8892124&5263&1.01&0.72&2&25.46\\
		4816098&6239&1.29&0.95&1&36.06&8916436&6566&1.35&0.87&1&32.29\\
		4850965&5503&1.04&0.61&2&20.67&9146690&5387&1.11&0.72&2&25.46\\
		4949214&6511&1.36&0.92&2&34.64&9206726&6876&1.31&0.46&4&14.50\\
		4949350&6587&1.40&0.88&2&32.75&9306290&5571&1.04&0.82&2&29.97\\
		4949766&6587&1.39&0.81&2&29.52&9393015&5877&1.01&0.24&2&6.41\\
		5038288&5785&0.99&0.88&2&32.75&9456932&5875&0.97&0.53&2&17.33\\
		5107198&6077&1.36&0.36&2&10.66&9474101&5945&1.10&0.21&2&5.42\\
		5273178&6774&1.32&0.88&2&32.75&9594038&6694&1.31&0.94&4&35.58\\
		5397765&6251&1.34&0.94&2&35.58&9640204&6620&1.33&0.53&2&17.33\\
		5426665&6323&1.38&0.39&2&11.79&9640472&6076&1.34&0.34&2&9.92\\
		5444276&6475&1.31&0.71&2&25.01&9710612&5867&1.08&0.39&2&11.79\\
		5450307&6398&1.24&0.99&3&37.98&9730249&6479&1.34&0.91&2&34.16\\
		5480545&6535&1.31&0.93&2&35.11&9896552&6279&1.26&0.87&1&32.29\\
		5514866&5487&0.97&0.28&2&7.77&9897710&5840&1.08&0.43&2&13.32\\
		5514871&5220&1.06&0.28&2&7.77&9965888&5589&1.13&0.31&2&8.83\\
		5543840&6518&1.20&0.82&2&29.97&9970838&6429&1.25&0.96&2&36.54\\
		5623538&6729&1.32&0.99&1&37.98&10023062&6469&1.38&0.89&2&33.22\\
		5623852&5886&1.10&0.57&2&18.98&10134084&5926&1.00&0.55&5&18.15\\
		5629449&6897&1.31&0.71&1&25.01&10490282&5504&1.05&0.79&2&28.60\\
		5646176&6302&1.20&0.99&1&37.98&10614890&5283&1.06&1.00&2&38.46\\
		5795235&6517&1.36&0.91&2&34.16&10809099&6051&1.31&0.91&2&34.16\\
		5898014&6697&1.35&0.83&2&30.43&11017401&5648&1.09&0.80&2&29.06\\
		5988566&6299&1.20&0.44&2&13.71&11018874&6454&1.30&0.99&2&37.98\\
		6114118&6234&1.24&0.94&2&35.58&11247377&6184&1.38&0.40&2&12.17\\
		6114140&6384&1.16&0.93&3&35.11&11349677&6076&1.23&0.84&1&30.90\\
		6145032&6315&1.28&0.81&1&29.52&11400413&6781&1.34&0.76&4&27.25\\
		6149358&6660&1.28&0.89&2&33.22&11498689&5464&1.10&0.31&2&8.83\\
		6219870&5663&1.05&0.81&1&29.52&11653059&6160&1.26&0.29&2&8.12\\
		6224148&6230&1.18&0.20&2&5.09&11924842&5494&1.13&0.84&5&30.90\\
		6385867&5306&1.06&0.58&1&19.40&11969131&6444&1.23&0.63&1&21.53\\
		6386598&6658&1.37&0.76&2&27.25&12067121&6211&1.33&0.43&5&13.32\\
		6391602&5782&0.99&0.42&2&12.94&12108612&5695&1.09&0.71&2&25.01\\
		6421219&6191&1.36&0.79&2&28.60&12119534&5296&0.98&0.64&2&21.96\\
		6449077&6366&1.31&0.94&2&35.58&12121738&6134&1.31&0.73&2&25.90\\
		6529902&6604&1.38&0.29&2&8.12&12157161&6513&1.26&0.78&2&28.15\\
		6693864&6846&1.35&0.86&1&31.82&12157799&6117&1.17&0.89&5&33.22\\
		6836589&5628&1.15&0.73&2&25.90&12354328&5251&0.97&0.81&2&29.52\\
		6846595&6718&1.26&0.99&1&37.98&12356839&5605&1.14&0.35&2&10.29\\
		6854461&6547&1.39&0.95&3&36.06&12418959&6427&1.36&0.78&2&28.15\\
	\end{xltabular}
	\begin{minipage}{14cm}
		\small \textbf{Notes:} Effective temperature $T_{\rm  eff}$ and radius $R_{\odot}$ was taken from (\textit{Gaia}-DR2).\\
		\textbf{References:} (1) \citet{Santos2021}, (2) \citet{McQuillan2014}, (3) \citet{Reinhold2015}, (4) \citet{Chowdhury2018}, (5) \citet{Yang2019}.\\\\
	\end{minipage}
	
\begin{figure}[h!]
		\centering
		\includegraphics[width=0.6\textwidth]{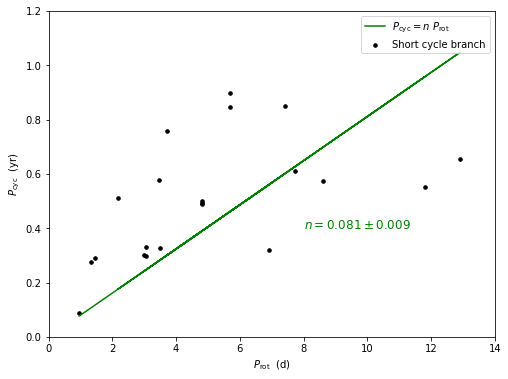}
		\caption{$P_{\rm rot}$ vs. $P_{\rm cyc}$ using a simple linear regression without an intercept for a sample of stars whose $P_{\rm cyc}$ is less than 1 year. The determined fit of $P_{\rm rot}$ vs. $P_{\rm cyc}$ relation are shown as a solid green line.}
		\label{simple-linear-fit}
	\end{figure}

	After predicting the values of the activity cycles for our extended,  compared to \citet{Mittag2023}, data sample, we wish to examine the theoretical prediction given by equation 2 on short $P_{\rm cyc} <$ 1 yr. 
	This is because the latter equation is a theoretical prediction, based on first physical principles,
	as opposed to empirical fit, which lacks any theoretical or conceptual justification.
	Therefore, we focused on the activity cycles derived from previous studies, as presented in Table 1. We chose 20 stars whose $P_{\rm cyc}$ is less than a year and plot the fit between $P_{\rm rot}$ and $P_{\rm cyc}$ as shown in Figure \ref{simple-linear-fit} using a simple linear regression without an intercept given by
	\begin{equation}
		P_{\rm cyc}\;[{\rm yr}]= n \ P_{\rm rot}\;[{\rm d}].
		\label{eq7}
	\end{equation} 
	We obtained the slope $n$= 0.081 $\pm$ 0.009 and $R^{2}$ value is 0.997. {Although the $R^{2}$ value for the fit is near unity, there is a large scatter indicating a poor quality fit.}
	Note that $P_{\rm cyc}$ here is in years, as in Figure 14 from \citet{Mittag2019}.
	Therefore, for the lower and upper bounds of our
	138 Kepler IDs with $P_{\rm rot}$ ranging from 0.202 d to 0.997 d,
	this simple theoretically justified equation predicts for 
	$P_{\rm cyc}=0.081\times0.202\times365.25=5.98$ d and $0.081\times0.997\times365.25=29.50$ d,
	which are not very different from applying the more accurate powerlaw fit using equation \ref{eq6} of
	{5.09 d and 38.46 d}, respectively.
	
	\section{Conclusions}\label{section:conclusion3}
	In this work, we studied the empirical relation between 
	star activity cycle and rotation period. 
	First, we reproduced the fit between $P_{\rm rot}$ and $P_{\rm cyc}$ using \citet{Mittag2023} data
	and obtained the following fit parameters {$\log P_{\rm cyc} \approx (1.488 \pm 0.092) + (1.324 \pm 0.067) \log P_{\rm rot}$, which are the same parameters as \citet{Mittag2023}.}
	Then, using a larger data set made up of  $P_{\rm rot}$ and their  associated $P_{\rm cyc}$ taken from prior studies, we again re-examined the fit between $P_{\rm rot}$ and $P_{\rm cyc}$ and obtained the followinh fit parameters 
{$\log P_{\rm cyc}\approx (1.585 \pm 0.064)+(1.256 \pm 0.051) \log P_{\rm rot}$.} 
Using these new parameters, we applied this relation to a sample of 83 F-type and 55 G-type main sequence stars whose rotation periods of less than one day. {The objective was to predict short activity cycles for these stars, aiming to ascertain in future studies if short activity cycles are a common occurrence in these stars or not.} 
As a result we derived 138 predicted $P_{\rm cyc}$ ranging from {5.09 d to 38.46 d} , which are listed in Table \ref{tab:less than 1d}.

Usefulness of measuring short stellar activity cycles 
hinges on two main general difficulties:

(i) If monitoring program of stellar activity (e.g. activity-related chromospheric emission S-index or similar)  is used 
as in references such as \citet{Mittag2019}; or \citet{Baum2022}, then cadence time of observations is too long 
e.g. according to table 2 from the latter reference cadence could be 87 observations per year i.e. 365/87 = 4 days. Resolving activity cycles with {$5.09\leq P_{\rm cyc}\leq 38.46$} d with such cadence would be nearly impossible.

(ii) If Kepler data light curves are used for e.g. plotting number of flares per day vs. time then large number of flare-detection would be necessary to have a reliable statistics. However, the problem is long cadence, 30 minutes, for the mainstream Kepler data. The photometer used by Kepler is sensitive to wavelengths ranging from 400 to 865 nm, covering the entire visible spectrum and a fraction of the infrared. The accuracy of the photometer of Kepler is approximately 0.01\% or 0.1 mmag, when 30-minute integration times are used while considering stars with a magnitude of 12. Kepler's 30-minute integration detected flare amplitudes less than 0.1\% of the stellar value and energies of $2\times10^{33}$ ergs. The duration of the flares ranged from one to three hours, with a rapid increase followed by a slow, exponential decline \citet{Maehara2012}. When Kepler data is taken at a higher cadence or sampling rate of one minute, the accuracy of the measurements decreases. However, this higher cadence enables Kepler to detect flares that are too brief to be detected reliably using the main 30-minute integrations. With the one-minute cadence, Kepler can detect flares with energies as low as $10^{32}$ ergs \citet{Maehara2015}.

It is worth noting that earlier studies exist using different observations where the energy involved in the observed transient brightening is estimated to range from $10^{25}$ to $10^{29}$ erg \citet{Shimizu1995}. Also, as far as the Sun is concerned, studies exist \citet{Mason2023} which consider flare frequency as a function of flare energy in the range $10^{27}$to $10^{31}$ erg, but this is applicable to the Sun only.

In order to have a good statistics for Kepler IDs considered, we need to detect flares with energies $10^{27-32}$ ergs in order to see variation number of flares per day on a time scale of {$5.09\leq P_{\rm cyc}\leq 38.46$} d. 
To achieve this goal a new space mission is necessary with short time cadence ($< 1$ minutes) and photometric accuracy $< 0.01$\%. 

{A typical example of such a proposed hypothetical space mission would record data of the number of flares per day for each target. 
These data can be presented in bins of e.g.  one-day width 
where the bin heights would show the number of flares detected in that bin. These bins would then show a periodic variation over time. 
Fitting a sinusoidal curve then would enable to deduce activity cycle
period. Thus, through this periodic variation, we could potentially 
detect the target's magnetic activity cycle period.
In some sense our approach is similar to \citet{Scoggins2019}. However,
their observation was so short in duration that only decrease
in the flare activity was seen. A longer duration
observations from a proposed new space mission would enable to see periodic variation and hence deduce the activity cycle period. 
}

Alternative option could be making more short cadence ground-based s-index  monitoring program of stellar activity with cadence $\approx 1$ d or less. However it is unclear 
whether this is technically feasible.
In any case, the present study provides predictions for $5.09\leq P_{\rm cyc}\leq 38.46$ d and
we hope that future either space or ground-based observational missions will put to test our predictions.
Unitl such time the jury is still out.

\section*{Acknowledgements}
Some of the data presented in this paper were obtained from the Mikulski Archive for Space Telescopes (MAST). STScI is operated by the Association of Universities for Research in Astronomy, Inc., under NASA contract NAS5-26555. Support for MAST for non-HST data is provided by the NASA Office of Space Science via grant NNX13AC07G and by other grants and contracts.\\
Authors would like to thank Deborah Kenny of STScI for kind assistance in obtaining the data, Cozmin Timis and Alex Owen of Queen Mary University of London for the assistance in data handling at the Astronomy Unit.\\
A. K. Althukair wishes to thank Princess Nourah Bint Abdulrahman University, Riyadh, Saudi Arabia and  
Royal Embassy of Saudi Arabia Cultural Bureau in London, UK for the financial support of her PhD scholarship, held at Queen Mary University of London.\\
{Authors would like to thank an anonymous referee whose comments greatly improved this manuscript.}

\section*{Data Availability}
Some of the data underlying this article were accessed from Mikulski Archive for Space Telescopes (MAST) \url{https://mast.stsci.edu/portal/Mashup/Clients/Mast/Portal.html}. This paper also has made use of data from the European Space Agency (ESA) mission {\it Gaia} (\url{https://www.cosmos.esa.int/gaia}), processed by the {\it Gaia} Data Processing and Analysis Consortium (DPAC, \url{https://www.cosmos.esa.int/web/gaia/dpac/consortium}). Funding for the DPAC has been provided by national institutions, in particular the institutions participating in the {\it Gaia} Multilateral Agreement. The derived data generated in this research will be shared on reasonable request to the corresponding author.

\bibliography{RAA-2023-0222.R1}{}
\bibliographystyle{aasjournal}

\end{document}